\documentclass[pdflatex,sn-mathphys, iicol]{sn-jnl}

\usepackage[capitalize]{cleveref}
\usepackage{xcolor}
\usepackage{csquotes}
\usepackage{braket}

\jyear{2022}%

\newcommand{\captiontitle}[1]{{\bf{}#1}}

\newcommand{\dd}{\mathrm{d}}

\newcommand{\bvec}[1]{\boldsymbol #1}

\definecolor{p-channel}{RGB}{0,153,230}
\definecolor{d-channel}{RGB}{40,37,110}
\definecolor{c-channel}{RGB}{242,56,20}

\makeatletter
\def\convertto#1#2{\strip@pt\dimexpr #2*65536/\number\dimexpr 1#1}
\makeatother
\newcommand{\measurewidths}{{%
  \color{red}
  :::
  TEXT \convertto{in}{\the\textwidth} in
  :::
  COLUMN \convertto{in}{\the\columnwidth} in
  :::
  HEIGHT(ex) \convertto{pt}{1ex} pt
  :::
  WIDTH(em) \convertto{pt}{1em} pt
  :::
}}

\begin{document}

\title[Integrators for FRG]{Better Integrators for Functional Renormalization
Group Calculations}

\author[1,2,3]{\fnm{Jacob} \sur{Beyer}}\email{beyer@physik.rwth-aachen.de}
\equalcont{These authors contributed equally to this work.}

\author[1]{\fnm{Florian} \sur{Goth}}
           \email{florian.goth@physik.uni-wuerzburg.de}
           \equalcont{These authors contributed equally to this work.}

\author*[1]{\fnm{Tobias} \sur{M\"uller}}
           \email{tobias.mueller@physik.uni-wuerzburg.de}
           \equalcont{These authors contributed equally to this work.}

\affil[1]{\orgdiv{Institute for Theoretical Physics},
          \orgname{University of W\"urzburg},
          \orgaddress{\city{W\"urzburg}, \postcode{97074}, \country{Germany}}}

\affil[2]{\orgdiv{Institute for Theoretical Solid State Physics},
          \orgname{RWTH Aachen University},
          \orgaddress{\city{Aachen}, \postcode{52056}, \country{Germany}}}

\affil[3]{\orgdiv{School of Physics},
          \orgname{University of Melbourne},
          \orgaddress{\city{Parkville}, \postcode{VIC 3010}, \country{Australia}}}

\abstract{
    We analyze a variety of integration schemes for the momentum space
    functional renormalization group calculation with the goal of finding an
    optimized scheme.
    Using the square lattice $t-t'$ Hubbard model as a testbed we define and
    benchmark the quality.
    Most notably we define an error estimate of the solution for the ordinary
    differential equation circumventing the issues introduced by the divergences
    at the end of the FRG flow.
    Using this measure to control for accuracy we find a threefold reduction in
    number of required integration steps achievable by choice of integrator.
    We herewith publish a set of recommended choices for the functional
    renormalization group, shown to decrease the computational cost for FRG
    calculations and representing a valuable basis for further investigations.
}

\maketitle

\section{Introduction}
\label{sec:Introduction}

The functional renormalization group (FRG) has proven itself to be a versatile
theoretical framework to study competing ordering tendencies and other many-body
phenomena both in itinerant
fermionic~\cite{salmhofer_honerkamp_2001_theory,metzner_salmhofer_honerkamp_2012,platt_hanke_thomale_2013}
and quantum spin
systems~\cite{ReutherOrig,reuther2014cluster,Iqbal3D,BuessenDiamond,PyrochlorePRX}. The underlying
flow equations are a large system of coupled, non-linear ordinary
differential equations (ODEs), the solution of which stipulated the development
of highly sophisticated numerical codebases~\cite{tagliavini-ea-2019-multiloop,
Buessen2021, kiese2020multiloop}.
So far, algorithmical improvements mainly focussed on the efficient
representation of vertex functions and optimizing momentum integrations on the
right-hand-side (RHS) of the flow equations, with the numerical solution of the
ODE itself only treated as a neccessity rather than an opportunity for
improvements. Consequently, a simple Euler step was used in the majority of FRG
calculations to date, see e.g. \cite{platt_hanke_thomale_2013, ReutherOrig,
Buessen2021}, with the inception of higher-order solvers from the Runge-Kutta
family being a fairly recent development in the
field~\cite{kiese2020multiloop,thoenniss2020multiloop,bonetti2021single,Markhof_2018,Weidinger_2017,PhysRevB.89.045128,Hauck_2021}.

The appropriate choice of integrator will lead to improved accuracy at the same
number of evaluations of the RHS, which constitutes the main computational
bottleneck of any FRG calculation.
Alternatively, one can significantly reduce this number while
maintaining good accuracy. In contrast to many other optimizations, this gain
does not introduce any further physical approximations, but solely rests in
mathematics.
As a prototypical setup to study the influence of different integrators, we
choose the momentum space FRG in the simplest truncation scheme, only
considering the four-particle vertex $\Gamma^{(4)}$ while neglecting self-energy
effects. Additionally, we limit ourselves to the well-understood case of the
square lattice Hubbard model with hopping up to second nearest
neighbors~\cite{honerkamp-salmhofer-2001,honerkamp_doping_2001,honerkamp_salmhofer_furukawa_rice,honerkamp2003,honerkamp_interaction_2004,honerkamp_ferromagnetism_2004,salmhofer_renormalization_2004,husesalm,Uebelacker-2021-sffeedback,husemann_frequency_2012,LichtensteinPHD,tagliavini-ea-2019-multiloop,hille-ea-2020-quantitative,hille_pseudogap_2020,bonetti2021single,schaefer-wentzell-multi-2021}.

We explicitly refrain from implementing more sophisticated approximation
schemes~\cite{vilardi2017nonseparable,hille-ea-2020-quantitative,tagliavini-ea-2019-multiloop,reckling-honerkamp-2018,Uebelacker-2021-sffeedback,bonetti2021single} or multi-band extensions~\cite{thomale2009functional,kiesel-thomale-ea-2012,wang-frg-graphene-2012,platt_hanke_thomale_2013,lichtenstein-boeri-ea-2014,klebl2021moire},
not only for clarity of analysis, but rather focus on benchmarking a large
variety of different standard ODE solvers, most available in existing
libraries~\cite{rackauckas2017differentialequations,boost_odeint}. Nevertheless we expect our
conclusions to provide a starting ground for the application of better integrators to
more sophisticated physical approximations.

The paper is organized as follows: After an introduction of our notation of
momentum space FRG and the approximations employed in our implementation in
\cref{ssec:frg_theory}, we provide an overview over various integration schemes
in \cref{ssec:integrators} and define tangible metrics to judge the quality of
the different schemes in \cref{ssec:metrics}. In \cref{sec:results} we employ a
two-stage elimination procedure. In a first step we sort out all integrators
incapable of reproducing the physical ordering instability for the
nearest-neighbor only model. We subsequently analyze the remaining set over a
larger parameter space including second-neighbor hopping to distinguish optimal
choices.

\section{Methods}
\label{sec:method}

\subsection{FRG in brevity}
\label{ssec:frg_theory}
The following discussion of FRG will be reduced to an absolute minimum and only
serves as a means to introduce the notation used. For detailed
derivations we refer the interested reader to standard
literature~\cite{salmhofer_honerkamp_2001_theory, metzner_salmhofer_honerkamp_2012,
platt_hanke_thomale_2013}.

In the following, we will focus solely on the square-lattice $t-t'$ Hubbard
model defined by
\begin{align}
    \begin{split}
        H_0 =&
            -\! \sum\limits_{\braket{i,j},\sigma} t \, c_{i,\sigma}^\dagger
            c_{j,\sigma}^{\phantom\dagger} \,
            -\! \sum_{\braket{\braket{i,j}},\sigma} t' \, c_{i,\sigma}^\dagger
            c_{j,\sigma}^{\phantom\dagger} \\
            & -\! \sum_{i,\sigma} \mu\, c_{i,\sigma}^\dagger
            c_{i,\sigma}^{\phantom\dagger}\,,
            \label{eqn:hubbard_h0}     
        \end{split}  \\
        H_\mathrm{I} =& U\, \sum_{i,\sigma} \, n_{i,\sigma} n_{i,\bar\sigma} \,,
            \label{eqn:hubbard_hI}
\end{align}

where $t$ ($t'$) denote (next-) nearest-neighbor hopping amplitudes, $\mu$ is
the chemical potential and $U$ the strength of the onsite Hubbard interaction.
Due to the spin-rotation invariance of \cref{eqn:hubbard_h0} and
\cref{eqn:hubbard_hI}, we can constrain our derivation to the
$\operatorname{SU}(2)$-symmetric set of equations.  To simplify the functional
dependence of the FRG equations, we invoke the static approximation of the
problem, i.e. consider only the zero Matsubara frequency component of the
vertex functions, and neglect self-energy effects
\cite{honerkamp2003,platt_hanke_thomale_2013}.

\begin{figure}
    \centering
    \includegraphics[width=\columnwidth]{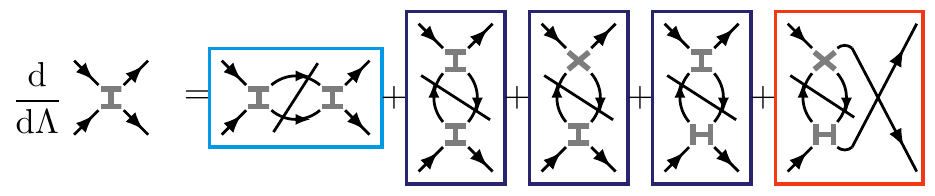}
    \caption{\captiontitle{Diagrammatic representation of the functional
    renormalization group equations.} We show the  $\operatorname{SU}(2)$-symmetric
    flow equation, where the $\operatorname{SU}(2)$-degree of freedom is
    kept constant along the lines of the vertex. The line through the loop
    represents the scale derivative $\dd/\dd\Lambda$. We have indicated the
    three channels of the FRG flow, differing in the
    bosonic transfer momentum: \textcolor{p-channel}{$P$}-,
    \textcolor{c-channel}{$C$}-, and \textcolor{d-channel}{$D$}-channel.}
    \label{fig:diags}
\end{figure}

The $\operatorname{SU}(2)$-reduced FRG equation in the conventional truncation
scheme including contributions up to the four-point vertex $\Gamma^{(4)}$ is
diagrammatically shown in \cref{fig:diags} and can be expanded to (using $V$ as
$\operatorname{SU}(2)$-symmetrized four-point vertex):
\begin{align}
    \frac{\dd}{\dd \Lambda} &V_{\bvec k_0, \bvec k_1, \bvec k_2} = \nonumber \\
    % P-Channel
    & \phantom{+} \,\,\, V_{\bvec k_0, \bvec k_1, \bvec l}
        \dot L^{-,\Lambda}_{\bvec l, -\bvec l + \bvec k_0 + \bvec k_1}
        V_{\bvec l, -\bvec l + \bvec k_0 + \bvec k_1, \bvec k_2} \nonumber \\
    % D-Channel 1
    &+  V_{\bvec k_0, \bvec l-\bvec k_0 + \bvec k_2, \bvec k_2}
        \dot L^{+,\Lambda}_{\bvec l, \bvec l - \bvec k_0 + \bvec k_2}
        V_{\bvec l, \bvec k_1, \bvec l - \bvec k_0 + \bvec k_2} \nonumber \\
    % D-Channel 2
        \label{eqn:flow}
    &+  V_{\bvec k_0, \bvec l - \bvec k_0 + \bvec k_2, \bvec k_2}
        \dot L^{+,\Lambda}_{\bvec l, \bvec l - \bvec k_0 + \bvec k_2}
        V_{\bvec k_1, \bvec l, \bvec l - \bvec k_0 + \bvec k_2} \\
    % D-Channel 2
    &+  V_{\bvec k_0, \bvec l - \bvec k_0 + \bvec k_2, \bvec l}
        \dot L^{+,\Lambda}_{\bvec l, \bvec l - \bvec k_0 + \bvec k_2}
        V_{\bvec l, \bvec k_1, \bvec l - \bvec k_0 + \bvec k_2} \nonumber \\
    % C-Channel
    &+  V_{\bvec k_0, \bvec l-\bvec k_1 + \bvec k_2, \bvec l}
        \dot L^{+,\Lambda}_{\bvec l, \bvec l + \bvec k_1 - \bvec k_2}
        V_{\bvec l, \bvec k_1, \bvec k_2} \nonumber \, ,
\end{align}
where we suppress the, in the $\operatorname{SU}(2)$ Hubbard model, unnecessary
spin indices.
We also use a modified Einstein sum convention to indicate the
momentum-integration in $\bvec l$ over the Brillouin zone (BZ).
To treat the continuous momentum-dependence of the vertex function, we employ
the \enquote{grid-FRG} momentum discretization scheme in the BZ with the
additional refinement-scheme described in Ref.~\cite{beyer_hauck_klebl}.
\cref{eqn:flow} will commonly be referred to as the right-hand side (RHS) of the
FRG equation, which
we need to compute at each flow step iteration.

The $\dot L^{\Lambda}$ given in the RHS equation is the derivative w.r.t. the
scale $\Lambda$ of the regulated loop
 \begin{equation}
 L^{\pm,\Lambda} = \sum\limits_n  G^\Lambda(\bvec k,\mathrm{i}\omega_n) G^\Lambda(\bvec k',\pm \mathrm{i}\omega_n)\, .
 \end{equation}
The FRG regulator $\Theta(\Lambda)$ here is introduced multiplicatively by defining $G^\Lambda = \Theta(\Lambda)G$ where $G$ is the bare propagator.
We have used both the particle-particle loop $L^{-, \Lambda}$ and the
particle-hole loop $L^{+, \Lambda}$ in out notation.
As a regulator we choose the $\Omega$-cutoff $\Theta(\Lambda)
= \frac{\omega^2}{\omega^2-\Lambda^2}$~\cite{husesalm} with high numerical stability.
The analytic Matsubara frequency summations then yield:
\newcommand{\ebl}{\epsilon} %
\newcommand{\eblp}{\epsilon'} %
\begin{align}
    \dot L^{\pm, \Lambda}_{\bvec l, \bvec l'} &=
        \frac{\dd}{\dd \Lambda}L^{\pm, \Lambda}_{\bvec l, \bvec l'}\\
     &=
    \begin{cases}
      \displaystyle
    \frac{\pm 1/4}{\ebl \mp \eblp}
    \left(
    \frac{\ebl(3\lvert\ebl\rvert + \Lambda)}{(\lvert\ebl\rvert + \Lambda)^3}
    \mp\frac{\eblp(3\lvert\eblp\rvert+\Lambda)}{(\lvert\eblp\rvert+\Lambda)^3}
      \right) \\[1em]
      \displaystyle
    \mp\frac{3\ebl^2-4\lvert\ebl\rvert\Lambda
    -\Lambda^2}{4(\lvert\ebl\rvert+\Lambda)^4}
    \qquad \text{if } \ebl = \eblp \ .
    \end{cases}
    \label{eqn:Ldot_omega}
\end{align}
To obtain the effective low-energy vertex we integrate the differential equation
(RHS) starting at infinite (i.e. large compared to bandwidth)
$\Lambda=\Lambda_\infty$ and approach $\Lambda \rightarrow 0$.
When encountering a phase transition the loop
derivatives will diverge in conjunction with the associated susceptibilities
and we terminate the flow. This is the integration for which we attempt to find
an optimized solver.
Due to the divergence we can not finalize the calculations using the solvers
but instead hard-terminate them once the maximum element of the vertex exceeds a
threshold $V_{\max}$.

\subsection{Integrators}
\label{ssec:integrators}
The possible choices of integrators are plentiful but in the following we highlight some
of the most prevalent algorithms as well as some which will prove to excel during
our testing.
The measurement schemes used to determine quality are provided at the end of
this section.
In order to numerically solve the non-linear non-autonomous differential equation
\begin{equation}
    \frac{\dd}{\dd\Lambda} V = \mathrm{RHS}(V, \Lambda)
 \label{eq:frgsimplified}
\end{equation}
we utilize both single-step and multi-step methods \cite{Hairer2008} in order to obtain
an iteration procedure
\begin{equation}
 V_{n+1} = V_n + \Delta_{\Lambda,n} \Phi(\Lambda_n, V_n, \Delta_{\Lambda,n})
\end{equation}
that we terminate at the divergence of $V$.

We almost exclusively focus on integrators that are implemented in the excellent
\texttt{DifferentialEquations.jl} package
\cite{rackauckas2017differentialequations} written in the Julia programming
language.  We emphasize integrators that are well-known and commonly employed,
such as the Runge-Kutta class and also widely available in other programming
languages.
A full list of the considered (including all disregarded) integrators can be found in
\cref{sec:A_integrators}.

\subsubsection{Single-Step Methods}
A single-step method is characterized by utilizing at each step a starting value
$V_n$ and additional evaluations of the RHS to construct $V_{n+1}$. The most
well-known family of methods in this class are the Runge-Kutta type integrators where the
function $\Phi$ is constructed as a weighted average of evaluations of RHS
within the interval $[\Lambda_n, \Lambda_{n+1}]$ such that it coincides up to
the respective order with its Taylor polynomial:
\begin{eqnarray}
 V_{n+1} = V_n + \Delta_{\Lambda}\sum_{i=1}^s b_i k_i\\
    k_i = \mathrm{RHS}(\Lambda_n + c_i \Delta_{\Lambda},
        V_n + \Delta_{\Lambda} \sum_{j=1}^s a_{ij} k_j)
\end{eqnarray}
The method is explicit if $a_{ij}=0$ for $j \geq i$ else it is an implicit
method. Adaptivity can be included by a time-step dependent $\Delta_{\Lambda,n}$.
The specifics of the methods are covered in tremendous detail in the
literature, e.g., \cite{Hairer2008}.

\subsubsection{Example: Adaptive Explicit Euler}
The conceptually simplest integrator in FRG that serves as the baseline for this work
is the adaptive explicit Euler described in \cref{alg:euler}. Note that we have
included a function $f(\Lambda, V_{\max}) =
\min(\max(a\Lambda/V_{\max}, \Delta_{\min}), \Delta_{\max})$
which is an adaption scheme specifically designed to reduce the step-width in
the proximity of the expected flow divergence.
Here $a$ is a small number determining the relative speed of the integration.

\begin{algorithm}
\caption{Explicit adaptive Euler}\label{alg:euler}
\begin{algorithmic}[1]
\State $V = V_0$, $\Lambda = \Lambda_0$
\While{$\max(V) < V_{\max}$ and $\Lambda > 0$}
    \State $\dd V = \mathrm{RHS}(V_{n}, \Lambda)$
    \State $V_{n+1} = V_n + \Delta_{\Lambda} \dd V$
    \State $\Lambda_{n+1} = \Lambda_n - \Delta_{\Lambda}$
    \State $\Delta_{\Lambda} = f(\Lambda, V_{\max})$
\EndWhile
\end{algorithmic}
\end{algorithm}

\subsubsection{A note on implicit Runge-Kutta methods}
We have attempted to include implicit integration schemes in our analysis but were
inhibited by their high memory cost. While for explicit methods
the requirements are understood to be $n\times\texttt{sizeof}(V),
n\in\mathbb{N}$, for the implicit methods we must instead consider the size of
the Jacobian we are attempting to invert.
This is proportional to the square of the size of a single solution (as
it is a linear map between two of them) and is thus
$\propto\texttt{sizeof}(V)^2$.
Even for the relatively small scale systems we employ here the number of
elements is $n_{\bvec k}^6 \approx 10^{13}$. \emph{All} implicit integrators
will therefore not be a viable option for FRG calculations.

\subsubsection{Methods based on the Non-linear Magnus Series}
Another class of one-step methods specifically suited for homogeneous,
non-linear, non-autonomous ODEs is based on suitable approximations $\Omega^{[m+1]}$ of the Magnus operator $\Omega$ \cite{Casas2006} in the formal exact
solution of
\eqref{eq:frgsimplified},
\begin{equation}
 V(\Lambda) = \exp \left(\Omega(\Lambda, V_\infty) \right) V_\infty
 \label{eq:magnussolution}
\end{equation}
where $V_\infty = V(\Lambda = \infty)$.
For the following approximations we define the matrix $A$ by the relation
$\mathrm{RHS}(\Lambda, V) = A(\Lambda, V) V$.
In this overview of methods we restrict ourselves to the approximations of first,
\begin{equation}
 \Omega^{[1]}(\Lambda, \Lambda_\infty) = \int_{\Lambda_\infty}^\Lambda A(s,V_\infty) ds
 \label{eq:magnus1}
\end{equation}
and second order
\begin{equation}
 \Omega^{[2]}(\Lambda, \Lambda_\infty) = \int_{\Lambda_\infty}^\Lambda A(s, e^{\Omega^{[1]}} V_\infty) ds.
 \label{eq:magnus2}
\end{equation}
and refer the reader to the literature \cite{Casas2006} for expressions
of the higher order in terms of iterated commutators.

An iterative procedure is obtained by applying the approximate time evolution operator
$\exp(\Omega^{[m]})$ for small steps $\Delta_{\Lambda}$ such that
\begin{equation}
 y_{n+1} = \exp(\Omega^{[m]}) y_n.
\end{equation}
 Approximating the integrals in
\cref{eq:magnus1} and \cref{eq:magnus2} by order-consistent low-order
approximations, such that e.g., at first order
\begin{eqnarray*}
 \Omega^{[1]}(\Lambda+\Delta_\Lambda, \Lambda) & = \int_{\Lambda}^{\Lambda+\Delta_\Lambda} A(s,V(\Lambda)) ds\\
 &\approx \Delta_\Lambda A(\Lambda, V(\Lambda)) ds.
\end{eqnarray*}
Adaptivity is easily included by using time slices of differing times
$\Delta_{\Lambda,n}$ determined by the same strategy as for the adaptive Euler.

\subsubsection{Multistep/Adams methods}
In contrast to one-step methods, multistep methods utilize
previous evaluations of the right hand side, $f_{n+k} = \mathrm{RHS}(V_{n+k},
\Lambda_{n+k})$ in order to approximate it with a Lagrange polynomial.
The general form of these linear multistep methods can be stated as
\begin{equation}
    \sum_{j=0}^k \alpha_j V_{n+j}
        = h \sum_{j=0}^k \beta_j \mathrm{RHS}(\Lambda_{n+j}, V_{n+j}).
\end{equation}
If $\beta_k = 0$ the method is explicit and does not require the evaluation of
RHS at unknown points. Commonly this class is termed Adams-Bashforth
technique.  Otherwise the method is implicit. If the interpolation polynomial is
only augmented with the single yet unknown point $\mathrm{RHS}(\Lambda_{n+k},
V_{n+k})$ the method is termed Adams-Moulton method. In this case the implicit
nature can be treated by employing a predictor-corrector scheme where the
prediction $\hat{V}$ is obtained by an explicit Adams method of one order lower
than the implicit method, RHS is evaluated and the new value $V_{n+k}$ is
obtained.
Multistep methods need startup values, but these are obtained with explicit
Runge-Kutta or similar methods. Adaptive step size integrators can be obtained
by utilizing more sophisticated interpolation techniques.

\subsection{Measures of Quality}
\label{ssec:metrics}
Differentiation of the quality of integrators will be based on the following
aspects:
\begin{enumerate}
    \item Error compared to a converged high-fidelity Euler calculation
    \item Minimal number of RHS-evaluations
    \item RAM requirements
\end{enumerate}

\subsubsection{Error compared to high fidelity Euler}
All integrators must converge to the same result at infinite numbers of steps
performed, or equally at negligible integration error for adaptive procedures.
To obtain this result in the most controlled fashion we converge an
adaptive Euler integrator to a high number of steps
($N_{\mathrm{RHS}} \approx 6000$) as a reference.
The measure of error employed in the comparison is the normed sum of differences
in the resulting vertex tensor:
\begin{equation}
    \frac{\sum_i \lvert V(i) - V_{\mathrm{Euler}}(i)\rvert}
    {\sum_i\lvert V_{\mathrm{Euler}}(i)\rvert} \,.
\end{equation}
To avoid numerical discrepancies in the vicinity of the flow divergence, we instead compare the vertices slightly above the critical
scale $\Lambda_{\mathrm{crit}}$ at $\Lambda=0.17$.
This scale is sufficiently low for the integrator to have performed a significant number of
integration steps but sufficiently high to avoid the critical scale. Note
that this number is arbitrarily chosen to lie within this region.

Because this measure does not include an analysis of the error size which is
tolerable for qualitatively correct FRG results (for the given approximation
level), we have additionally in stage 2 \cref{ssec:stage2} included a full phase
scan to ensure the results are consistent with expected behavior. We eliminate
all integrators that are inconsistent.

\subsubsection{Minimal number of RHS-evaluation}
As we are aiming to optimize the calculation time requirements, the most
important measure of quality for any integration routine must be the number of
RHS-evaluations it requires until divergence.
This is the only expensive part of the calculation and is thus a trivial, but
platform independent estimator for the runtime.
While this is no longer true for implicit integrators where the inversion of
the Jacobian would be most expensive, these are disallowed by their RAM
requirements and thus can be neglected here.

\subsubsection{Memory requirement}
For most FRG calculations the bottleneck is not only the calculation time but
also the memory consumption
\cite{beyer_hauck_klebl,lichtenstein_high-performance_2017}.
As higher-order methods may require the saving of intermediate results we want
to track the number of concurrently allocated vertex-sized objects as a measure
of RAM usage.
Once again for implicit methods this may not be the most accurate measure due to
the creation of high-dimensional Jacobian matrices, but we disregard them
entirely.
The impact on memory usage by other objects is of second order when compared to
the vertices in scaling:
The vertices scale as $N_{\bvec k}^3$ while the loop-derivatives scale as
$N_{\bvec k}^2$ and the dispersion cache scales as $N_{\bvec k}N_{\bvec k_f}$.
The measure chosen is the peak memory usage during the calculations. This will
be highly proportional to the total number of vertex objects though slight
lower-order effects are to be expected.

\section{Results}
\label{sec:results}
We have ensured that our implementation falls within the FRG equivalence class
of the Hubbard model reported on in Ref.~\cite{beyer_hauck_klebl}.
This binary equivalence is sufficient as a proof of correctness.
Note that while one of the codes in that publication is written by one of us,
we utilize for this benchmark an independent implementation written in the Julia programming language.

All simulations were performed using the following parameters:
$t=1, U=3t, t'\in [0.0,..., -0.5] t$, $\mu = -4t'$.
We use a momentum meshing of $n_{\bvec k} = 16\times16$ with a refined meshing
of $n_{\mathrm{fine}}=9\times9$.
The FRG flow is started at $\Lambda_\infty=50$ and tracked until the maximum element of
the vertex exceeds $V_{\max}=50t$, where we terminate the integration and
analyze the last vertex $V_{\Lambda_{\mathrm{crit}}}$ with respect to its ordering tendencies.
While this resolution is insufficient for physical simulations at low scales and
we acknowledge that the stability of all integrators increases for
higher resolutions, the fact that most integrators calculate the proper results
is testament to a sufficient resolution for this analysis.

We have split the investigation into a two staged elimination process where we
iterate over consecutively more involved tests to find the best set of
integrators and neglect the remainder.

\subsection{Stage 1}
The first approach to determine the feasibility of the integrators is to analyze
the square lattice Hubbard model at $t'=0$, $\mu=0$ where we expect
antiferromagnetic ordering.

\begin{figure}[h]%
    \includegraphics[width=\columnwidth]{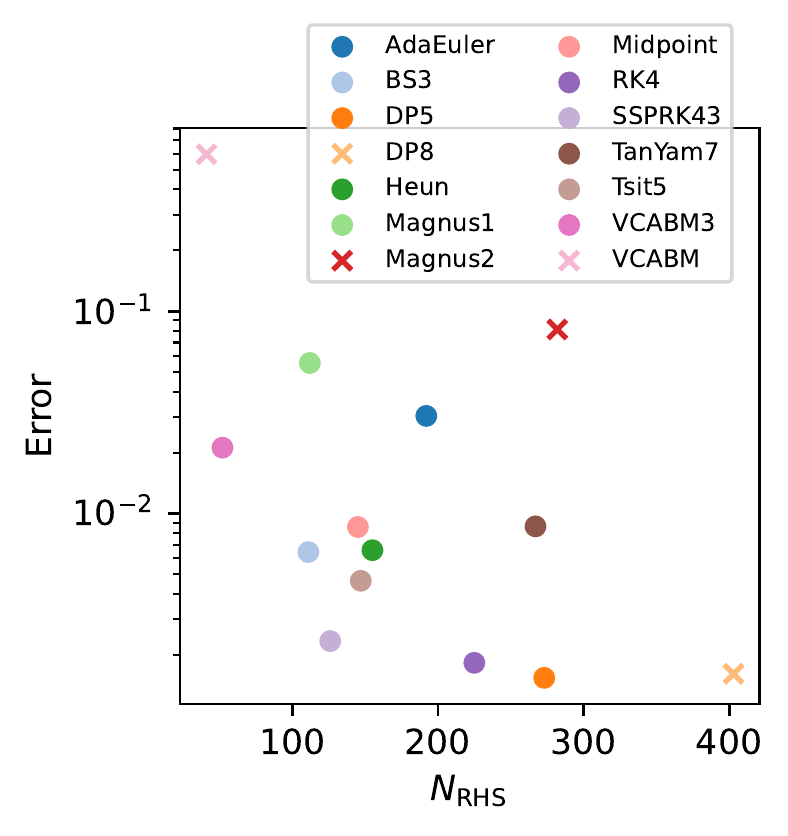}
    \caption{\captiontitle{Integrator Comparisons}
    Quick analysis of the quality of integrators. For this measurement we chose
    the single point at $t'=0$, $\mu=0$ in the $t-t'$ square-lattice Hubbard
    model. At this point we expect an antiferromagnetic divergence, every method
    incapable of producing this is omitted.
    We show the number of steps required by the integrators for the prediction
    of the divergence as well as the error when compared to high-fidelity Euler
    calculations.
    The best integrators lie near the origin of the coordinate system, low
    number of steps and negligible error. We in the following will consider only
    integrators within the marked region, the set of which we will refer to as
    stage 2 integrators. A full list of considered integrators as well as the
    reasons some are not shown can be found in \cref{sec:A_integrators}.}
\label{fig:integrator}
\end{figure}

The results of stage 1 are displayed in \cref{fig:integrator}.
We now select the subset of integrators which yield a speedup (or are similar) in
number of evaluations when compared to the adaptive Euler integration scheme
while maintaining higher or similar accuracy.
This reduces our list of integrators to consider for the continuation to be:

\begin{itemize}
    \item Adaptive Euler
    \item Bogacki-Shampine 3/2 (BS3)
    \item Dormand-Prince's 5/4 Runge-Kutta (DP5)
    \item Second order Heun's
    \item Second order Midpoint
    \item canonical Runge-Kutta 4 (RK4)
    \item Strong-stability preserving Runge-Kutta (SSPRK43)
    \item Tsitouras 5/4 Runge-Kutta (Tsit5)
    \item Third order Adams-Moulton, BS3 for starting values (VCABM3)
    \item Tanaka-Yamashita 7 Runge-Kutta (TanYam7)
\end{itemize}

We have thus significantly reduced the number of integrators to consider in the
next section.
While we acknowledge that some of these might have been tweaked
into compliance by optimizing parameters we insist on the out-of-the-box setup
of the integrators being at least sufficient (if maybe not optimal).

\subsection{Stage 2}
\label{ssec:stage2}
To check the reduced set of integrators for physical consistency we now
evaluate the $t-t'$ square lattice Hubbard model at van Hove filling by scanning
the second neighbor hopping $t'$ (adapting the chemical potential $\mu=4t'$
accordingly to remain at v.H. filling) as previously calculated in
Refs.~\cite{LichtensteinPHD, salmhofer_honerkamp_2001_theory}.
We perform these calculations with each of the second stage integrators to obtain a
more qualified understanding of their accuracy and cost.

\begin{figure}[h]%
    \centering
    \includegraphics[width=\columnwidth]{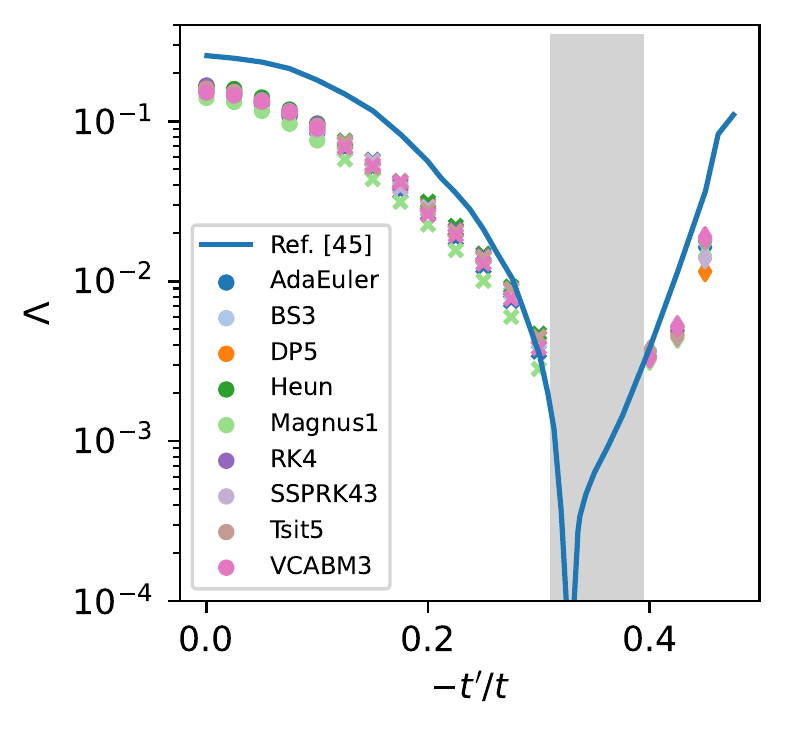}
    \caption{\captiontitle{Hubbard Model} Calculations of the reference Hubbard
    model at van Hove filling with $t'\in[0.0, -0.4t]$.
    The comparative data is taken from \cite{Honerkamp_2001}, on which we have superimposed the subset of
    stage 2 integrators. We have removed all integrators who were unable to
    produce the correct phase at points. Note that we in this analysis ignore
    the region around the SC-FM phase transition as it is notoriously unstable
    under low scales \cite{LichtensteinPHD,beyer_hauck_klebl}.
    Dots represent antiferromagnetism, crosses superconductivity and diamonds
    ferromagnetism.}
\label{fig:hubbard}
\end{figure}

To determine accuracy we in a first step check that the phase transitions occur
in a controlled manner in all integrators and all points are properly found 
to diverge. In this we however make an exception near the SC-FM phase
transitions, where the low integration resolutions used here \emph{will} lead to
uncontrolled behavior.
By this process we eliminate the Midpoint and TanYam7 integration methods, which
did not properly reproduce the leading instability for some points in the phase
diagram. The remaining results and integrators are shown in \cref{fig:hubbard}/
We can here also see the good agreement of qualitative results from the
different integrators, an expected but satisfying result.

\begin{figure}[h]%
    \includegraphics[width=\columnwidth]{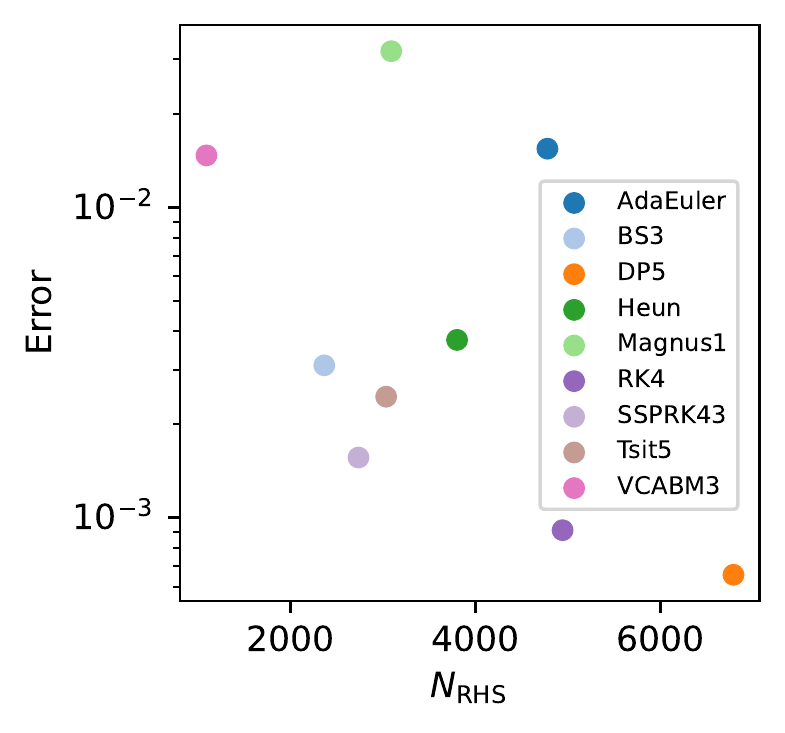}
    \caption{\captiontitle{Stage 2 Integrator Comparisons}
    We compare the integrators which passed the primary analysis in a second
    iteration using an extended testing scheme. This extended scheme consists of
    a full $t-t'$ Hubbard model phase scan. The $N_{\mathrm{RHS}}$ here
    represents the total number of steps for the evaluation of the 20 points in
    the phase diagram.
    Similarly the error represents the mean of the errors at the four evaluation
    points $t'=0, -0.05, -0.2, -0.4$.
    Once again proximity to the origin is the quality measure in the graphic.
    }
    \label{fig:integrator_stage2}
\end{figure}

To evaluate the performance we sum the number of RHS evaluation for each of the 20 runs in the interval $t' \in [0,..., -0.5]$ into a total number $N_\mathrm{RHS}$.
This is a more accurate measure of performance than before as the phase diagram
contains parameter regions, where early divergences are expected as well as points, at which low
RG scales have to be reached.
This is supplemented by an improved error estimate: As before we calculate the
effective vertex at the low scale of $\Lambda = 0.17$ and compare it to the
results obtained by a converged high-fidelity Euler, but we now consider 3 additional
points, one in each phase at $t' = 0.0, -0.05, -0.2, -0.45$. As an accuracy
measure we consider the mean of the relative error for these four parameter
combinations.
The results of this analysis are presented in \cref{fig:integrator_stage2}.
We find, that depending on the exact requirements on the integrator, different
algorithms should be employed. For purely optimizing the runtime, i.e. the
number of RHS evaluations, the multi-step method VCABM3 clearly is the method of
choice, with a mean error comparable to the adaptive Euler, but only needing
less than a quarter of RHS evaluations. On the other extreme, we find DP5 to be
the most accurate, but requiring 40\% more runtime. A competitive alternative is
RK4 with a slightly larger relative deviation from the reference data, but at a
computational cost comparable to the Adaptive Euler. As a good compromise
between numerical complexity and accuracy, we find a cluster of three methods:
BS3, Tsit5 and, most notably, SSPRK43 all have an order of magnitude lower
relative error, but at roughly two third of the computational cost of the
Adaptive Euler.

\begin{figure}[h]%
    \includegraphics[width=\columnwidth]{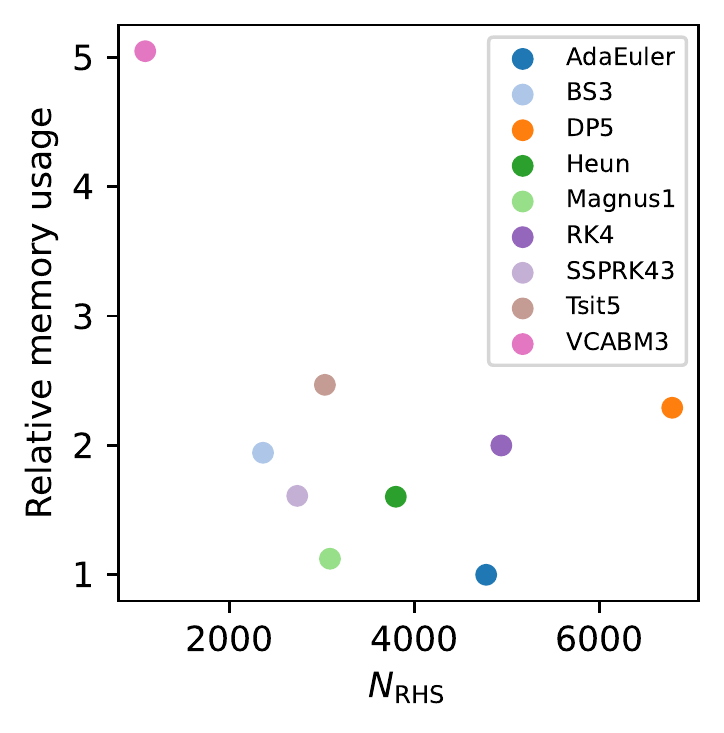}
    \caption{\captiontitle{Stage 2 Memory Comparison}
    Supplementing the analysis presented in \cref{fig:integrator_stage2} we here
    provide the measurements of the memory usage during the integration
    procedure.
    We have denoted the memory usage in terms of the most efficient schemes, the
    adaptive Euler. Note that we have just measured the total memory usage,
    partial might thus correspond to objects of smaller size.
    Nonetheless the measurement will reflect the ram needed for the integrations
    very well.
    }
    \label{fig:ram_stage2}
\end{figure}

In \cref{fig:ram_stage2} we additionally show the measured peak memory
requirement of our implementation relative to the Adaptive Euler. This figure is
highly implementation dependent, but should nevertheless serve as a rough guide
to the overhead incurred by the different solvers. Most clearly, this figure
shows, that the extreme numerical efficiency of VCABM3 is achieved trading
runtime for memory consumption, with a fivefold requirement compared to a simple
Euler implementation. The remaining integrators considered all fall in the range
of 1.5 to 2.5 the memory requirement. Out of the three general algorithms,
SSPRK43 clearly has the lowest memory requirements, which combined with its good
accuracy and numerical efficiency makes it the best suited ODE integrator in our
benchmark, followed closely by BS3.

\section{Conclusions}\label{sec13}

We have benchmarked a multitude of ODE integration algorithms both from the
\texttt{DifferentialEquations.jl} library and own implementations against a
reference Adaptive Euler using the square-lattice Hubbard
model as a test case. We have identified, that the right choice of integrator
can significantly reduce the numerical cost for integrating the runtime while at
the same time producing more accurate results. We identify the multi-step
algorithm VCABM3 to be the most efficient one with acceptable numerical errors,
while highest accuracy can be achieved by the numerically expensive DP5
algorithm, or alternatively the slightly less accurate RK4, which however has
about the same numerical cost as the Adaptive Euler.

As the best compromise between accuracy and numerical speedup, we identify
Tsit5, BS3 and SSPRK43, with the latter slightly outclassing its rivals.  This
fact at first glance is surprising, as this specific integrator is designed to
handle dissipative differential equations stemming from a discretization of
hyperbolic partial differential equations. However, as the FRG equations are
believed to flow towards a fixed point for any given phase, we speculate them to
exhibit dissipation in the mathematical sense, which is the property SSPRK43 is
design to utilize. This means that solutions of the RG flow corresponding to
different initial conditions in the same phase become more similar as they
approach $\Lambda=0$, as the dominant physics will be the same for all of these.
A mathematically thorough analysis of the FRG equations regarding this property,
however, is beyond the scope of this paper.

As we only have analyzed a subset of the plethora of available ODE integrators
in this paper, we cannot claim to have found the \enquote{best} integrator for
FRG system, but we ensured to represent a spectrum of the common choices.

We also have focused on the simplest physically interesting model, the
square-lattice Hubbard model. While we believe, that the scaling of
computational cost can be extrapolated to other problems, more complicated
problems posed with more intricate Fermi surfaces may
require denser integration steps, the best algorithms may therefore differ.

Furthermore when using other approximations of momentum space FRG, for example
self energy inclusions or multiloop, the RHS equations are subject to structural
changes.
The inclusion of self energies might benefit from the use of operator splitting methods,
such as the generalized Strang or Leapfrog splitting \cite{McLachlan1995},
where self energy and vertex are evaluated at alternating half-step intervals.
The merit of our results is still valid, the best integrators as found here
will be good candidates for the other applications.

Another path of algorithmic progress can be made by taking into account to
intricacies of the FRG equations. Since their integration starts at infinity we
propose to
investigate transformations of this semi-infinite domain as in
Ref.~\cite{macdonald,PhysRevB.89.045128,in_prep_hauck}.

Furthermore, the integration of the FRG equations contains an inherent
singularity corresponding to the physical ordering tendency. Implicit
methods~\cite{wanner1996} are well suited for such problems, but were out of
the scope of our investigation due to memory requirements.
It could be a worthwhile direction of future investigation.
More specialized multistep integrators that utilize rational interpolants~\cite{Karimov2021, wanner1996}
 instead of Newton and Lagrange polynomials
for the definition of their integration rule could be beneficial in dealing with
the singular behavior. But of course their effectiveness rests on methods
of obtaining the action of the Jacobian of $RHS$ or suitable approximations to it.

\backmatter

\bmhead{Acknowledgments}
J.B.thanks GRF (German Research Foundatation) for support through RTG
1995 within the Priority Program SPP 244 \enquote{2DMP}.
F.G. and T.M. thank the GRF for funding through the SFB 1170 “Tocotronics” under the grant
numbers Z03 and B04, respectively.
Furthermore we want to thank Jonas Hauck, Dominik Kiese, Lennart Klebl, Janik Potten, Stephan Rachel and
Tilman Schwemmer for helpful discussions.

\bmhead{Author's contributions}
J.B. and T.M. were leading the development of the code with F.G. taking charge
of the simulation effort. Analysis was conducted on a joint basis,
the manuscript was a joint effort by the authors. Overall no distinction can be
drawn in level of contribution between the authors.

\bmhead{Data availability}
The FRG implementation used in this paper is available at \url{https://doi.org/10.5281/zenodo.6391339}.
The numerical data shown is available from the corresponding author upon reasonable request.

\begin{appendices}

\section{Full list of considered integrators}
\label{sec:A_integrators}
\begin{figure*}
    \centering
\begin{tabular}{cp{3cm}p{8cm}}
    \textbf{Name} & Comment & Reason if discarded \\
    \hline
    AB3, AB5, ABM32, ABM54 & \cite{Hairer2008} & (Not adaptive)\\
    Adaptive Euler & \cite{Hairer2008} &\\
    Adaptive Magnus1 & \cite{Casas2006} & \\
    Adaptive Magnus2 & \cite{Casas2006} & (Error S1)\\
    BS3 & \cite{Bogacki1989} &\\
    CVODE & \cite{rackauckas2017differentialequations} & (complex arithmetic)\\
    DP5 & \cite{Dormand1980} &\\
    DP8 (Runtime S1) & \cite{Hairer2008} & \\
    EPIRK4s3A & \cite{rackauckas2017differentialequations} & (not adaptive)\\
    exp4 & \cite{rackauckas2017differentialequations} & (not adaptive)\\
    exprb32 & \cite{rackauckas2017differentialequations}\\
    Feagin10 (Wrong S1) & \cite{feagin2012high} & (issues calculating jacobian)\\
    FRK65 & \cite{rackauckas2017differentialequations} & (very excessive runtime)\\
    Heun & \cite{Hairer2008} &\\
    HochOst4 & \cite{rackauckas2017differentialequations} & (not adaptive)\\
    implicit Euler & \cite{wanner1996} & (Memory)\\
    implicit Midpoint &  \cite{wanner1996} & (Memory)\\
    KenCarp4 & \cite{rackauckas2017differentialequations} & (Memory)\\
    Kvaerno5 & \cite{rackauckas2017differentialequations} & (Memory)\\
    LawsonEuler & \cite{rackauckas2017differentialequations} & (Memory)\\
    Midpoint & \cite{Hairer2008} &  (Wrong Hubbard)\\
    NorsettEuler & \cite{rackauckas2017differentialequations} & (Not adaptive)\\
    QBD, QNDF, QBDF1, QNDF1 & \cite{rackauckas2017differentialequations} & (Memory)\\
    RadauIIa3, RadauIIa5 & \cite{rackauckas2017differentialequations} & (complex arithmetic)\\
    RK4 & \cite{Hairer2008} &\\
    RKO65 & \cite{rackauckas2017differentialequations} & (Not adaptive)\\
    ROCK2, ROCK4 & \cite{rackauckas2017differentialequations} & (excessive runtime)\\
    Rodas4p2 & \cite{rackauckas2017differentialequations} & (Memory)\\
    Ros3p & \cite{rackauckas2017differentialequations} & (Memory)\\
    Rosenbrock23 & \cite{rackauckas2017differentialequations} & (Memory)\\
    SDIRK & \cite{rackauckas2017differentialequations} & (Memory)\\
    SSPRK22, SSPRK33, SSPRK83 & \cite{rackauckas2017differentialequations} & (Not adaptive)\\
    SSPRK43 & \cite{Kraaijevanger_1991} &\\
    SSPRK932 & \cite{rackauckas2017differentialequations} & (type errors, excessive runtime)\\
    TanYam7 & \cite{tanaka1992} & (Wrong Hubbard)\\
    Trapezoid & \cite{wanner1996} & (Memory)\\
    TRBDF2 & \cite{rackauckas2017differentialequations} & (Memory)\\
    Tsit5 & \cite{tsitouras2011} &\\
    VCAB3 & \cite{Hairer2008} &\\
    VCAB5 & \cite{Hairer2008} & (Error S1)\\
    VCABM3 & \cite{Hairer2008} &\\
    VCABM5 & \cite{Hairer2008} & (Error S1)\\
    veldd4, velds4 & \cite{rackauckas2017differentialequations} &(Memory)\\
\end{tabular}
    \caption{\captiontitle{Full set of considered integrators.} The disregarded
    integrators have the reason given for disregarding them.}
\end{figure*}

\end{appendices}

\bibliography{references}

\end{document}